\documentclass[a4paper]{jpconf}
\usepackage{graphicx}
\usepackage[utf8]{inputenc}
\usepackage{amssymb}
\usepackage{amsmath}
\usepackage{mathtools}
\usepackage[a4paper, total={6in, 8in}]{geometry}
\usepackage{latexsym}
\usepackage{geometry}
\usepackage{array}
\usepackage{amsmath,graphicx,epsfig}
\usepackage{epstopdf}
\usepackage{euscript}
\usepackage{amsfonts}
\usepackage{float}
\usepackage{tabularx}
\usepackage{bm}
\usepackage{upgreek}
\usepackage{mathrsfs}
\usepackage{breqn}
\usepackage{listings}
\usepackage{xcolor}

\newcolumntype{M}[1]{>{\centering\arraybackslash}m{#1}}
\newcolumntype{N}{@{}m{0pt}@{}}

\def\prn#1{{\left(#1\right)}}

\def\sbrk#1{{\left[#1\right]}}
\def\abrk#1{{\left\langle#1\right\rangle}}

\begin{document}
\title{Multi-messenger astronomy in the new physics modality with GPS constellation}

\author{Arko P. Sen$^1$}
\author{Kalia Pfeffer$^1$}
\author{Paul Ries$^2$}
\author{Geoffrey Blewitt$^1$}
\author{Andrei Derevianko$^1$}

\address{$^1$University of Nevada, Reno, NV 89511, USA}
\address{$^2$Jet Propulsion Laboratory, California Institute of Technology, 4800 Oak Grove Dr., Pasadena, CA
}

\ead{andrei@unr.edu}

\begin{abstract}\\

We explore a novel, exotic physics, modality in multi-messenger astronomy.   We are interested in exotic fields emitted by the mergers and their  {\em direct}  detection with a network of atomic clocks. We specifically focus on the rubidium clocks onboard satellites of the Global Positioning System.  Bursts of exotic fields may be produced 
during the coalescence of  black hole  singularities, releasing quantum gravity messengers. To be detectable such fields must be ultralight and ultra-relativistic and we refer to them as exotic low-mass fields (ELFs). Since such fields possess non-zero mass, the ELF bursts lag behind the gravitational waves emitted by the very same merger. Then the gravitational wave observatories provide a detection trigger for the atomic clock networks searching for the feeble ELF signals.  ELFs  would imprint an  anti-chirp transient across the sensor network. ELFs can be detectable by atomic clocks if they cause variations in fundamental constants. We report our progress in the development of techniques to search for ELF bursts with clocks onboard GPS satellites. We 
focus on  the binary neutron star merger GW170817 of August 17, 2017. We find an intriguing excess in the clock noise post LIGO gravitational wave trigger. Potentially the excess noise could be explained away by the increased solar electron flux post LIGO trigger.
\end{abstract}

\section{Introduction}\label{Sec:Intro}
Since the initial discovery of gravitational waves (GW) by LIGO in 2015~\cite{LIGOfirstObservation2016}, there have been multiple detection of GWs. Most of these GWs result from mergers of black hole binaries.  However, on  August 17, 2017, a new class of GW  events was discovered: the binary neutron star merger GW170817~\cite{LIGOVirgo-NeutronStarMerger2017}.  That was the first astrophysical source detected in both the GW and multiwavelength electromagnetic radiation modalities and such is referred to as the multi-messenger astronomy~\cite{Abbott_multimessenger_2017}. The messengers came from a host galaxy 40 megaparsecs ($\sim$100 million light-years) away. Ref.~\cite{dailey2020ELF.Concept} extended the gravitational and electromagnetic modalities of multi-messenger astronomy to exotic, beyond the Standard Model of elementary particles, fields. We start by summarizing salient points of the concept paper~\cite{dailey2020ELF.Concept} and then report on our progress in the search for such exotica in Global Positioning System (GPS) atomic clock datastreams.

\begin{figure}
\center
\includegraphics[width=0.75\textwidth]{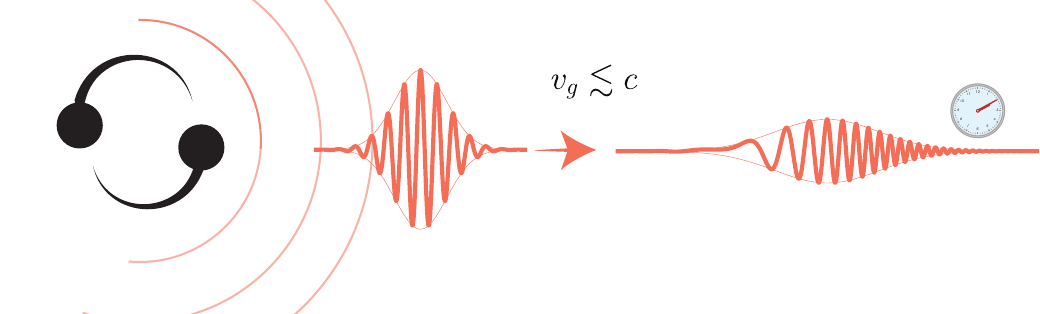}
\caption[]{ 
An emitted ELF burst propagates with  the group velocity $v_g \lesssim c$ to the detector, lagging behind the GW burst.  
Since the more energetic ELF components propagate faster, the arriving ELF wavepacket exhibits  a characteristic  frequency anti-chirp. From Ref.~\cite{dailey2020ELF.Concept}.
}
\label{Fig:Cartoon}
\end{figure}

Fig.~\ref{Fig:Cartoon} summarizes the idea of multi-messenger astronomy with the addition of exotica modality.  A merger  emits both the GW and an exotic  low-mass fields (ELFs) bursts. Since such fields possess non-zero mass, the burst propagates at a group velocity $v_g$ smaller than the speed of light. Thereby, the ELF pulse lags behind the GW burst. Because of the dispersion, the ELF burst tends to spread out as it travels. More energetic components of the  ELF wave-packet travel faster, imprinting a universal anti-chirp signature in the atomic clock data. We refer the reader to Ref.~\cite{dailey2020ELF.Concept} for technical details and to Ref.~\cite{Derevianko2023-Moriond.ELF} for a more qualitative and informal discussion.

As to the exotic physics modality, we focus on ultralight (yet non-zero mass) bosonic fields as the messengers.  We assume that  bursts of such ELFs could be generated by cataclysmic astrophysical events such as black hole or neutron star mergers~\cite{bini2017deviation,Baumann2019}. Considering that much of the underlying physics remains to be discovered, a pragmatic  approach~\cite{dailey2020ELF.Concept}  is based on energy arguments: based on the  uncertainties for the LIGO events,
  $\Delta E \sim M_\odot c^2$ of energy  can be released in the form of ELFs from a black hole binary and $\Delta E \sim 0.1 \,M_\odot c^2$ in a neutron star merger. These are extreme  (compared to terrestrial experiments) amounts of energy potentially released in the ELF modality.  Atomic clocks are sensitive to gentle perturbations of energy levels by coherent, classical waves. Due to copious numbers of emitted ELFs  this condition is indeed satisfied all the way to the sensors. Notice that the atomic clocks sensitivities are such that we do not need require $\Delta E \sim M_\odot c^2$ for the ELFs to be detectable. Much smaller fractions of the solar mass would suffice for LIGO detectable events.

There is a wide variety of speculative scenarios~\cite{dailey2020ELF.Concept} for coherent ELF production in the mergers. One particularly intriguing scenario is due to quantum gravity. Much of the underlying physics of coalescing singularities in black hole mergers remains unexplored as it requires understanding of the as yet unknown theory of quantum gravity~\cite{Loeb2018-BH-signularities}. Then the ELF burst would emerge from the merger as a  quantum gravity messenger. 

Relativistic massive fields satisfy the energy-momentum dispersion relation
 \begin{equation}
\omega(k)=\sqrt{(c k)^2+\Omega_c^2},
\label{Eq:dispersion}
\end{equation}
where the  Compton frequency $\Omega_c = mc^2 /\hbar$ depends on  the ELF mass $m$.  The energy and momentum of  ELF quanta are given by $\varepsilon = \hbar \omega$ and $p=\hbar k$, where $k$ is the wave number. Since we require a close time correlation (say delay less than a week) between the arriving GW and ELF bursts and the vast burst travel distances, the ELFs must be necessarily relativistic, i.e. their group velocity $v_g \approx c$.

A propagation of the initially emitted Gaussian ELF pulse (of central frequency $\omega_0$ and of pulse duration $\tau_0$) leads to  the  solution~\cite{dailey2020ELF.Concept} for a field at the clock a distance $R$ away from the progenitor  
\begin{align}
\phi(R,t) \approx & \frac{1}{R} \left(\frac{c\Delta E}{2\pi^{3/2} \omega_0^2 \tau}\right)^{1/2} \exp\left(  -\frac{(t-t_s)^2}{2 \tau^2} \right) \nonumber\\
                  & \times \cos\left( \omega_0 (t-t_s) - \frac{\omega_0}{4\delta t}(t-t_s)^2\right)\,,
\label{Eq:DispersionModeldE}
\end{align}
where $t_s = t_{GW} +  \delta t$  is the time of arrival of the center of the pulse at the sensor. It is offset by  $t_{GW} = R/c$, the time of arrival of the GW messenger. 
$\Delta E$ is the total energy released in the ELF channel.
The signal duration $\tau$ and GW-ELF time delay $\delta t$  are related as
\begin{equation*}
 \tau \approx  2  \delta t \,    /(\omega_0 \tau_0) \, .
 \label{Eq:duration}
\end{equation*}
Note that the instantaneous ELF frequency is time-dependent, $\omega(t)  = \prn{ 1 - (t-t_s) /(2\delta t)}\omega_0$, exhibiting a  frequency ``anti-chirp" at the sensor.   The slope of the anti-chirp is given by 
$$d\omega/dt = - 1/\prn{\tau \tau_0}=-\omega_0/\prn{2\delta t}\,.$$ 
This reflects a qualitative fact that the more energetic (or higher frequency) components have a higher phase velocity 
$\omega/k$.
The waveform~\eqref{Eq:DispersionModeldE} is shown in Fig.~\ref{Fig:Cartoon}.

A GPS clock sampling rate in our dataset is 1~Hz, see Sec.~\ref{Sec:GPS-data-processing}. This limits $\omega_0$ from above.  These  frequencies fix energies $\varepsilon_0 =\hbar \omega_0$  of detectable ELFs  to  below $10^{-14} \,  \mathrm{eV}$. 
Practically, we are limited to a duration of the pulse at the sensor to about a month; this limits $\omega_0$ from below. In fact, our GPS data set, see Sec.~\ref{Sec:Data-Analysis}, is effectively limited to $\sim$ a day, requiring that $\omega_0 \gtrsim 2\pi/\text{day}$,
i.e. above $\sim 10^{-19}~\rm{eV}$. Moreover, it is desirable to observe at least several oscillations, so the anti-chirp can be observed. To summarize, 
\begin{equation}
 10^{-19}\, \mathrm{eV} \lesssim \varepsilon_0 \lesssim  10^{-14} \,  \mathrm{eV} \,. \label{Eq:ELF-energy-range}
\end{equation}

Now we know how an ELF pulse looks like at a clock. How can we detect them? The most obvious way how the clock frequencies can be affected is through the variation of fundamental constants, discussed in the following section.

\section{Transient variation in fundamental constants}\label{Sec:Trans-Var}
ELFs can couple to atomic clocks through interactions that lead to the variation of fundamental constants.
We are interested in the phenomenological Lagrangian for the SM-ELFs interactions 
\begin{align}
   \mathcal{L} = \hbar c\prn{-\sum_f\Gamma_f  m_f c^2\bar{\psi}_f\psi_f + \frac{\Gamma_\alpha}{4} F_{\mu\nu}F^{\mu\nu} }\phi^2\,.\label{Eq:Lagrangian}
\end{align}
Here $\psi_f$ are the SM fermion fields (of masses $m_f$) and $F_{\mu \nu}$ is the electromagnetic Faraday tensor. $\Gamma_X$ are coupling constants (with $X=\alpha,m_f$) quantifying the strength of the quadratic interaction between the  scalar $\phi$ and SM fields. We focus on the interactions that are quadratic in the ELF field $\phi$ for a number of reasons, like the U(1) invariance for complex-valued fields and the fact that the current limits on the couplings being much weaker for the quadratic couplings than for the linear ones.

The interaction Lagrangian~(\ref{Eq:Lagrangian}) leads to the effective redefinition of fundamental constants (FCs):
fermion masses and the fine structure constant,
\begin{align}
&m^{\mathrm{eff}}_f(\textbf{r},t)=m_f\sbrk{1+\Gamma_f \hbar c \phi^2(\textbf{r},t)}\,\label{Eq:Fermion-mass-Redef},\\ 
    &\alpha^{\mathrm{eff}}(\textbf{r},t)=\alpha \sbrk{1+\Gamma_{\alpha}\hbar c\phi^2(\textbf{r},t)}\,.\label{Eq:Fine-Structure-Const-Redef}
\end{align}

The variations in FCs modify clock frequencies and, thereby, the apparent time as measured by the clock.
In general, the sensitivity of an oscillator resonance frequency $\nu$ to the variation in FC $X$ can be quantified by the coefficient
\begin{align}
    \kappa_X=\frac{\partial\, \log \nu}{\partial \log X}\,.
    \label{Eq:kappaX}
\end{align}
In atomic clocks, the clock frequency is determined by frequencies of both the atomic transition  and the local oscillator (LO), see Sec.~\ref{Sec:GPS-clocks}. The LOs in GPS clocks are quartz oscillators with their fundamental mode frequencies depending on FCs. We can introduce the LO sensitivity coefficient $\kappa_X^{\mathrm{LO}}$, as in Eq.~\eqref{Eq:kappaX}. Moreover, in GPS we compare the satellite and terrestrial clocks, so, in general, we have to operate in terms of differential sensitivity coefficients, $K_X$, that take sensitivities of individual clocks into account.

As the ELFs are ultra-relativistic, their travel time from satellites to terrestrial atomic clock is about 0.1~s,  an order of magnitude smaller than our sampling time interval, $\Delta_t=1$~s. Here we used the fact that the GPS constellation diameter $D_{\rm{GPS}} \approx 5\times 10^4$~km. Moreover, all the clocks would experience approximately the same value of the field if the ELF wavelength $\lambda \approx { 2\pi \hbar  c}/{\varepsilon_0} \gg D_{\rm{GPS}}$. For the range~\eqref{Eq:ELF-energy-range} of detectable ELF energies, the corresponding wavelengths map into the range  {$10^{5} \lesssim \lambda \lesssim 10^{10}$}~km. This means that for $\varepsilon_0 \lesssim 10^{-13} \, \mathrm{eV}$, $\lambda \gg 10^4$~km and  the entire GPS constellation measures nearly the same ELF phase and can be considered as a composite yet point-like quantum sensor. In the opposite limit of $ \lambda \ll 10^5 \, \mathrm{km}$,  field values at different satellite clocks differ and  if we were to take an average over all the clocks at a given time, the ELF signal would appear as an excess noise.

The fractional frequency excursion due to quadratic interaction~\eqref{Eq:Lagrangian} reads
\begin{align}
    s(t) = \frac{\delta\nu(t)}{\nu_{\mathrm{clock}}}=\Gamma_{\mathrm{eff}}\hbar c\phi(t)^2 \,,\label{Eq:ELFsig}
\end{align}
where  $\nu_{\mathrm{clock}}$ is the unperturbed or the nominal clock frequency, $\phi(t)$ is the ELF field at the sensor~\eqref{Eq:DispersionModeldE},  and $\Gamma_{\mathrm{eff}}\equiv \sum_X K_X\Gamma_X$ is the effective coupling constant expressed in terms of the differential sensitivity coefficients, $K_X$. 

The sensitivity coefficients $K_X$ depend on the internal composition and design of the clock and data taking rate. Our GPS clock phase (or bias) data, see Sec.~\ref{Sec:Data-Analysis}, is sampled  every second. These clock biases are the differences between the satellite  and the fixed reference terrestrial atomic clock times as measured by the clocks. We will focus on  $^{87}\mathrm{Rb}$ microwave GPS clocks and fix an active hydrogen maser located in Kourou, French Guiana (station ID: KOUR)  as a reference clock.  In the following section, we discuss the operation of such atomic clocks.

If the characteristic time scale of the transient variation of FCs is longer than the servo loop time constant, $K_X=\kappa_X^{\mathrm{atom}}$, otherwise $K_X=\kappa_X^{\mathrm{LO}}$. From Ref.~\cite{Campbell2021, FlambaumEtAl2004, Roberts2018a}, we have $\kappa_{\alpha}^{\mathrm{Rb}}\approx 4.34$, $\kappa_{m_e}^{\mathrm{Rb}}=2$, $\kappa_{m_q}^{\mathrm{Rb}}\approx -0.069$, $\kappa_{\alpha}^{\mathrm{H}}\approx 4$, $\kappa_{m_e}^{\mathrm{H}}=2$, $\kappa_{m_q}^{\mathrm{H}}\approx -0.150$, $\kappa_{\alpha}^{\mathrm{LO}}=2$ and $\kappa_{m_e}^{\mathrm{LO}}=3/2$, where $m_e$ and $m_q$ are the electron and average up and down quark masses respectively. As discussed in Sec.~\ref{Sec:GPS-data-processing}, our data is acquired in the regime when $K_X=\kappa_X^{\mathrm{atom}}$. Thereby, the effective coupling constant for the Rb-H pairwise comparison becomes
\begin{align}    \Gamma_{\mathrm{eff}}=0.34\,\Gamma_{\alpha}+0.081\,\Gamma_{m_q}\,, \label{Eq:GammaEff}
\end{align}
where the dependence on $\Gamma_{m_e}$ cancels out. Notice that the ELF signal would disappear in the data if we were to compare clocks of the same kind such as Rb-Rb, since $\Gamma_{\mathrm{eff}} \equiv 0$;  this is a powerful test of systematic effects. 

\section{Atomic clocks of GPS network}\label{Sec:GPS-clocks} 

GPS satellites host either $^{87}$Rb or $^{133}$Cs clocks. We focus on 
the Rb clocks because of their abundance (24 out of total 26 clocks are Rb in our data set for the GW170817 event) and because their frequency noise is primarily white~\cite{DanzeyRiley}.  The $^{87}$Rb $5S_{1/2}$ electronic ground state splits into a manifold of two, $F=1$ and $F=2$, hyperfine levels. The clock transition is between two $M_F=0$ B-field insensitive Zeeman sub-levels of the hyperfine manifold with the nominal frequency
$\nu_{\mathrm{Rb}}\approx 6.835 \, \mathrm{GHz}$. The frequency of a quartz oscillator is locked to $\nu_{\mathrm{Rb}}$ through a servo-loop operation, driving a frequency synthesizer to generate a 10.23 MHz digital clock. This then drives the rest of the electronics generating the GPS microwave signals which are transmitted on carrier waves precisely at $154 \times 10.23\rm{~MHz} = 1.57542 \rm{~GHz}$ (L1 band) and $120 \times 10.23\rm{~MHz} =
1.22760\rm{~GHz}$ (L2 band)~\cite{Blewitt2015307}. (Actually, the frequency of the digital clock is set slightly lower to allow for the mean combined relativistic effects of time dilation and gravitational red shift, so that the clock frequency is precisely 10.23 MHz according to an observer at the gravitational potential at mean sea level on Earth.)  The satellites broadcast microwave signals with a carrier phase that is commensurate with the on-board atomic clock system, through a phase locked loop frequency synthesizer. The phases of the microwave signals are further measured by the terrestrial receivers and through the data processing are interpreted as a measured GPS clock time, see Sec.~\ref{Sec:GPS-data-processing}. To estimate the effect of the FC variations, we need to gain a deeper understanding of how the GPS signal is generated, and how it propagates to the terrestrial receivers. We start with describing the operation of on-board Rb clocks in Sec.~\ref{Sec:Rb-atomic-clock} and then, in Sec.~\ref{Sec:H-Maser-atomic-clock}, the terrestrial H-masers.

\subsection{Rb atomic clocks}\label{Sec:Rb-atomic-clock}
The GPS Rb atomic clocks are vapour-cell type~\cite{Riley1981,Camparo}. Such clocks include several  packages which function together to ensure the frequency locking process. The Rb atoms are a part of the physics package (PP) which is primarily comprised of a  $^{87}$Rb discharge lamp, a filter cell containing $ ^{85}$Rb isotope, a resonance cell containing $^{87}$Rb, and a photo-detector~\cite{Riley1981}.  The resonance vapor cell is placed in a microwave cavity and a solenoid generating quantizing magnetic field~\cite{Camparo}. The PP is integrated with supporting electronic circuits which control temperature of the  discharge lamp, filter cell and resonance cell and also control the solenoid coil's current supply~\cite{Riley1981}.

Notice the use of two isotopes of Rb: $^{87}$Rb and $^{85}$Rb with nuclear spins $I=3/2$ and $I=5/2$ respectively. 
The $^{87}$Rb lamp emits on the transitions from the $5\, ^2\!P_{1/2},\,5\, ^2\!P_{3/2}$ levels to the ground state manifold (D1 and D2 lines).  The emitted light passes through a filter cell containing the $^{85}$Rb isotope. The fortuitous coincidence in the spectral line frequency associated with the $F=2$ hyperfine level of $^{87}$Rb with those associated with $F=3$ hyperfine level of $^{85}$Rb leads to the absorption of the $F=2$ spectral lines of D1 and D2 emissions of 
$^{87}$Rb from the discharge lamp by $^{85}$Rb in the filter cell~\cite{Guo}.

The filtered light is used for quantum state preparation via optical pumping to the $F=2$ clock state. When this $F=2$ state  is subjected to a microwave radiation, the microwaves drive the population between the  $F=2$ and $F=1$ clock levels. As a result the intensity of the light coming through the filter cell depends on the detuning of the microwave frequency from $\nu_{\mathrm{Rb}}$~\cite{Camparo,Guo}. The difference in light intensity is measured by the photodetector. This is used by a servo-loop circuit to generate an error signal which further controls the frequency of the LO that determines the microwave frequency. The magnitude of the error signal is the difference between the microwave frequency and $\nu_{\text{Rb}}$.

At the LO heart in GPS clocks are quartz crystals. Quartz is a piezo-electric material and  when an electrical field is applied across the crystal, it undergoes mechanical deformations and vibrates at a  fundamental frequency of the crystal that remains stable over relatively short timescales. The vibration modes depend on the geometry of the crystal. The relevant sensitivity coefficients to the variation of FCs was given in Sec.~\ref{Sec:Trans-Var}.

The servo-loop time constant
$\tau_{\mathrm{servo}}$ determines the duration of the fastest perturbation that can be recorded by the clock. Military operators manually adjust $\tau_{\mathrm{servo}}$ and its precise value is not available to us at a given epoch, however it is known to be within 0.01~s to 0.1~s~\cite{GriggsEtAl2015, Dupuis2008}. This is sufficient to suppress the effect of occasional frequency jumps in the quartz oscillator down to the level of $< 10^{-11}$~s in accumulated clock phase ~\cite{Riley1981}.  Our dataset, described in Sec.~\ref{Sec:ELF-search}, is sampled every second, giving enough time for LO frequency to be steered to the Rb atom frequency. This explains our choice of sensitivity constants in Eq.~\eqref{Eq:GammaEff}.  It is worth pointing out that the electronics itself can be affected by the varying FCs; we presume that all the derived frequencies follow the Rb atom frequency at a sufficiently fast time scale, so that the sampled phases of the L1 and L2 GPS signals by the terrestrial GPS receivers are a faithful representation of the behavior of the Rb atomic frequency.  

\subsection{H-maser atomic clock}\label{Sec:H-Maser-atomic-clock}
The terrestrial reference clock used for satellite clocks bias data (see Sec.~\ref{Sec:Data-Analysis}), is an active H-maser T4Science atomic clock housed in the station named KOUR, located in Kourou, French Guiana~\cite{IGS_KOUR,ITRF_KOUR}. Below  we focus on the general principles of H-maser operations~\cite{Major_Book}.  In these clocks, the quartz oscillator is locked to the transition frequency between  hyperfine levels attached to the ground state of the hydrogen atom, $\nu_{\mathrm{H}}\approx 1.420$~GHz. 

Such a clock uses a beam of atomic hydrogen,  obtained by dissociating of molecular hydrogen $\text{H}_2$ in a discharge chamber. The atomic beam passes through an inhomogeneous state-selecting magnetic field, which focuses hydrogens in the $F=1$ states into a Teflon-coated quartz storage bulb. This ``magic'' coating preserves atomic coherence for $\sim 10^6$ consecutive collisions with the bulb walls, critically enabling long interrogation times.

The bulb is placed within a  radiofrequency (RF) cavity  tuned to the transition frequency, $\nu_{\mathrm{H}}$ between $F=1,M_F=0$ and $F=0,M_F=0$ states of the H-atom~\cite{Klepner1962, Vessot_2005}. As the H-atoms undergo the transition from the $F=1$ state down to the $F=0$ state, the emitted microwave photons stimulate transitions in other atoms~\cite{Lombardi}, hence the maser qualifier in these clocks. {Cascade of stimulated transitions produce a highly coherent oscillation in the RF cavity. This results in a microwave signal to which the quartz oscillator is phase-locked.}

The RF cavity is enveloped by magnetic shields to attenuate the ambient fields. An inner solenoid coil creates a uniform magnetic field to separate the different magnetic sublevels.
There are two types of interactions that occur in the RF cavity: the atomic relaxation process that affects the population in $F=1$ state and the process that leads to the decay of atomic phase coherence that affect the frequency stability~\cite{Vessot_2005}. The major purpose of the bulb is to spatially confine the atoms, enabling  interaction with the in-phase components of the RF magnetic field in the cavity~\cite{Vessot_2005}.

For longer sustenance of the oscillation either a sufficiently high H-atom density is to be maintained or the RF cavity is to be interrogated with external RF radiation at $\nu_{\mathrm{H}}$. {Atomic clocks based on the former process is termed as active H-maser clocks, whereas clocks based on the latter are passive H-maser clocks. In active H-masers the cavity undergoes self-sustained oscillation}. Both  types of H-masers eventually drive a quartz oscillator which outputs a reference signal with a high stability and a low drift rate by control voltage adjustment~\cite{Hu}. The principal circuit of H-maser atomic clock has a crystal oscillator, an isolation amplifier circuit, an up and down-conversion circuit, a digital servo, and a frequency synthesizer circuit. There is an auxiliary circuit system that includes a constant temperature and a high-voltage source circuit, and a constant current source~\cite{Hu}. In passive H-maser clocks the interrogation microwave field is produced by an auxiliary frequency generator. H-masers have a good short-term stability,  $\sigma_y\prn{\tau=1~\mathrm{s}}<10^{-12}$, reaching a noise floor of~$\approx 10^{-15}$ after about an hour of averaging~\cite{Lombardi}.

The cavity frequency is stabilized by a servo loop.  The value of the time constant of the servo-loop response is in the order of 1~s~\cite{Audoin1981}. As with the Rb clocks, the response to a sudden change in FCs is that of a quartz for time-scales much shorter than the servo-loop time constant. For time scales $\gg$ than the servo-loop time constant, it is the response of the hydrogen hyperfine splitting that matters. Once again, we ignore the complicated response of electronic circuits to the FC variations, as we believe, such a response is only important for very short time scales.
These considerations govern our choice of sensitivity constants to FC variations in Sec.~\ref{Sec:Trans-Var}.

\section{GPS data processing}\label{Sec:GPS-data-processing}

As discussed in Sec.~\ref{Sec:GPS-clocks}, GPS satellites broadcast microwave signals with a carrier phase that is commensurate with the on-board atomic clock system, through a phase locked loop frequency synthesizer. The carrier phase is sampled at regular intervals by a global terrestrial network of GPS receiver stations.  Each GPS receiver measures the beat phase between the incoming carrier phase signal mixed with an internal replica signal, which is locked in phase to the receiver's internal quartz oscillator. Thus the beat phase measures the difference between time of transmission according to each GPS satellite clock, and the time of reception according to each receiver clock.  

The Jet Propulsion Laboratory (JPL) routinely analyzes carrier phase data every 30 seconds from the global GPS network~\cite{BERTIGER2020469}.  For the special purpose of this ELF search, JPL processed data at 1Hz across multiple days from an identical global 30-station network. Data processing includes modeling the time of flight of the carrier signal traveling at the phase velocity of light, accounting for relativistic Shapiro delay, rotational wind-up of circularly polarized phase, atmospheric refraction, and deterministic station motion caused by tidal and surface loading deformation. For each station, a linear combination of carrier phases at two different carrier frequencies is computed such that it eliminates the effect of ionospheric refraction $n_{\rm ion}$ assuming that $(n_{\rm ion}-1) \propto  1/\omega^2$~\cite{Blewitt_1989}. Note that this elimination applies even if fundamental constants are perturbed.  The linear combination conserves the satellite clock variations, which for GPS is designed to be identical on carrier signals of both frequencies.  Automatic data editing algorithms allow for arbitrary variation of clock phase, and so unusual clock behavior is not a determining factor in data outlier deletion~\cite{Blewitt_1990}.

Parameter biases in the model are simultaneously estimated using a square-root estimation filter.  Those parameters include all the satellite and receiver clock biases every second, with the exception of one high quality receiver clock which defines the reference coordinate time. The receiver reference clock is selected such that it is synchronized to a stable H-maser frequency standard. Other biases that are important to estimate but are not explicitly used in our subsequent analysis include station coordinates, satellite orbit parameters, atmospheric delays and their variation, and carrier phase biases associated with an ambiguity in the integer number of cycles, which are subsequently determined precisely and fixed~\cite{Blewitt_1989, Bertiger2010}. 

For the neutral atmosphere, the signal delay is non-dispersive at microwave frequencies, thus it is independent of frequency variations in the carrier signal that may be caused by FC variations.  However, in principle there could be a contribution of FC variations to delay arising from changes in refractive index.   Variations in refractive index due to weather are accounted for using the GPT2 model~\cite{Bohm-GPT2}, which was based on fits to numerical weather models of the European Center for Medium-Range Weather Forecasts.  

Any residual variations in refractive index, such those caused by lack of spatio-temporal resolution in the GPT2 model, are estimated as part of the multi-parameter bias estimation by the square root information filter.  Specifically, the variations are estimated as a random walk in the zenith delay and in two gradient parameters for each station.  These parameters are separable from satellite clock variations owing to the so-called ``mapping function," that is the modeled variation of atmospheric slant thickness with elevation angle of the satellite~\cite{Bohm-GPT2}.  Such estimation should also absorb variation in refractive index caused by the FC variations we are seeking, leaving the imprint of clock frequency variations intact.   The assumption is that any refractive index variations caused by ELFs do not greatly exceed the level of variations expected from weather.

The resulting data set used in our analysis is a 1-second time series of atomic clock biases from each of the GPS satellites that have Rb clocks. All of these biases are relative to the clock of the selected reference station.  We also have access to receiver clock biases, however most GPS stations only use quartz oscillators without an external atomic frequency standard, and so are of limited use.  Other useful metadata include satellite positions and velocities.

One aspect of the GPS system that we aim to exploit is that, in the co-rotating Earth-fixed reference frame of the global GPS station network, the satellite positions are designed to repeat approximately every sidereal day~\cite{Agnew-repeat,Larson-repeat}.   The precise repeat time must account for the precession of each satellite's orbit, caused by Earth's oblateness, and the orbit's eccentricity, inclination, and semi-major axis.  This precession is slightly different for each satellite, resulting in a repeat period of that varies in the range 1 day minus $245\pm5$s~\cite{Larson-repeat}.  This repeat of the orbit geometry causes systematic errors such as ground multipath reflections to also repeat, hence errors in the carrier phase measurement tend also to repeat.  Therefore, there is the potential to enhance signal to noise by comparing measured carrier phase time series from one repeat cycle to the next, as in the common mode error subtraction techniques.  Such methods have proven to greatly reduce scatter in geodetic GPS positioning every second, with accuracy at the few mm level~\cite{Larson-repeat}, therefore signal to noise ratio in clock bias data should similarly be improved.

\section{Search for ELF signal and preliminary results}\label{Sec:ELF-search}

\subsection{Effect of ELFs on the clock output}\label{Sec:ELF-clock-output}

The Rb-locked LO frequency is measured by a narrow band circuit whose signal can be related to a sinusoidal wave~\cite{0.A.Howe0.W.Allan1981}. The output voltage of such a circuit  can be expressed as
    \begin{align}
 V(t)=\sbrk{V_0+\varepsilon(t)}\sin\sbrk{2\pi \nu_{\mathrm{clock}}t+ \varphi (t)}\,,\label{Eq:Voltage}
    \end{align}
    where $V_0$ and $\nu_{\mathrm{clock}}$ are nominal peak voltage amplitude and the clock frequency respectively and $\varepsilon(t)$ and $\varphi(t)$ are the deviation from the nominal peak voltage and nominal clock phase~\cite{0.A.Howe0.W.Allan1981,Riley}. Usually the oscillator frequency remains constant for a short time interval  but it can drift over a relatively larger period of time. 

The instantaneous frequency is conventionally defined as the first derivative of the phase, which is 
    \begin{align}
              \nu(t) = \nu_{\mathrm{clock}}+\frac{1}{2\pi} \frac{d\varphi(t)}{dt}. \label{Eq:InstFreq}
    \end{align}
The extra phase term can be due to either technical noise or due to the ELF signal. Given a sampling time interval $\Delta_t$, the off-set phase is explicitly
    \begin{align}
     \varphi(t_j)&\nonumber = 2\pi \sbrk{ \int_{t_{j-1}}^{t_j} \nu(t')dt'-\nu_{\mathrm{clock}}\int_{t_{j-1}}^{t_j} dt'}\,\\
     &\approx 2\pi \sbrk{ \nu(t_j)- \nu_{\mathrm{clock}}}\Delta_t\,,\label{Eq:phase}
    \end{align}
where $t_j=j\Delta_t$ ; $j=\overline{1,N_t}$ and $N_t$ being to total number of data points.

Note that the deviation from the nominal clock frequency is $\delta \nu(t_j)=\nu(t_j)- \nu_{\mathrm{clock}}$. Then the fractional clock frequency excursions are
\begin{align}
   \frac{ \delta\nu(t_j)}{\nu_{\mathrm{clock}}}\approx \frac{1}{2\pi\Delta_t}\frac{\varphi(t_j)}{\nu_{\mathrm{clock}}}\,, \label{Eq:FracFreq}
\end{align}
to be matched to the sought ELF waveform~\eqref{Eq:ELFsig}.

\subsection{Data analysis and results}\label{Sec:Data-Analysis}
In our GPS.ELF group, we  pursue the analysis of data  from the GPS atomic clocks. One can think of the GPS constellation as the largest human-built $\sim 50,000 \,\text{km}$-aperture sensor array.  Combined with other satellite positioning constellations and terrestrial clocks, the total number of sensors $N_c \sim 100$.  Employing networks is crucial for distinguishing ELF signals from technical noise. Furthermore, by having baselines with the diameter of the Earth or larger, one can resolve the sky position of the ELF source. This is a critical feature for multi-messenger astronomy that enables correlation with other observations of the progenitor. However, an ELF feature (say leading edge) propagation time across the GPS constellation is $0.2 \, \text{s}$. The publicly available GPS data is sampled every 
$\Delta_t = 30\, \text{s}$. Our GPS.ELF group has developed techniques to generate more frequently sampled,  $\Delta_t = 1\, \text{s}$, GPS datastreams.
Still the limited time resolution makes tracking the leading edge of the ELF pulse across the GPS network challenging. 
Nevertheless, clock networks can still act collectively, gaining $\sqrt{N_c}$ in sensitivity and vetoing signals that do not affect all the sensors in the network.

We work with the GPS clock bias data for the day of GW170817 event detection and the day before. The satellite  clock biases are relative to a fixed reference terrestrial clock.
This is a non-stationary time series, dominated by random walk noise. This discrete clock bias data stream $d^{(0)}_{a,j}$, sampled every $\Delta_t=1$~s, can be expressed, for the $a^{\text{th}}$ satellite clock, in units of time as
\begin{align}
   {d^{(0)}_{a,j}}=\frac{\varphi_a(t_j)}{\omega_{a}}-\frac{\varphi_r(t_j)}{\omega_{r_0}}\,, \label{Eq:Clock-bias}
\end{align}
where the index $j=\overline{1,N_t}$, enumerates the epochs. $N_t$ is the number of points sampled during the orbital period, as discussed below. The frequencies $\omega_a$ and $\omega_{r_0}$ are the nominal clock angular frequency of a satellite $a$ and the reference terrestrial clocks. The time dependent discrete phases of the satellite and reference clocks are denoted by $\varphi_a(t_j)$ and $\varphi_r(t_j)$ respectively. The reference terrestrial clock for this work is a hydrogen maser at the KOUR station located in Kourou, French Guiana, see Sec.~\ref{Sec:H-Maser-atomic-clock}. 

Although the proposal paper~\cite{dailey2020ELF.Concept} has used the formalism of excess noise statistic and time-frequency decomposition, here we carry out a simplified analysis of the excess noise to also include a possibility of high-frequency/short wavelength ELFs, see Sec.~\ref{Sec:Intro}.

As discussed in Sec.~\ref{Sec:GPS-data-processing}, the orbital geometry of the satellites repeats relative to observer on Earth, leading to repeating multipath errors. To account for the multipath errors, the GPS satellite orbital period based on the Earth Centered Earth Fixed (ECEF) reference frame is set approximately to 24~hours minus 247~s~=~86,153~s. The sampling interval of 1~s thus leads to the total number of points per day $N_t =86,153$. 
To whiten the data, we employ the first order differencing of the discrete clock bias data, the so-called  discretized pseudo-frequency~\cite{Roberts2017-GPS-DM},
\begin{align}
       {d^{(1)}_{a,j}} = {d^{(0)}_{a,j+1}-d^{(0)}_{a,j}}\,,
\end{align}
where $a$ indexes individual datastreams and $j$ spans over sampling grid.
The differencing reduces the total number of discrete pseudo-frequency points to $N \equiv N_t-1$.
We remove the frequency drift and the offset for each clock $a$
\begin{align}
    u^{(1)}_{a,j}=d^{(1)}_{a,j}-f_{a,j}\,,
\end{align}
by fitting the ${d^{(1)}_{a,j}}$ data to the straight line of the form, $f_{a,j}=m_a t_j+b_a$, with $m_a$ and $b_a$ being fitting coefficients.  
The weighted average of $u^{(1)}_{a,j}$ over the network of $N_c$ satellite clocks is
\begin{align}
  \bar {d}^{(1)}_j= \frac{\sum_{a}^{N_c}  {u}^{(1)}_{a,j} /\sigma^2_a}{\sum_{a}^{N_c}1/\sigma^2_a}\,,\label{Eq:Network Wt Avg Pseudofreq}
\end{align}
where
\begin{align}
   &  \sigma_a^2=\frac{1}{N-1}\sum_{j=1}^N\prn{{u}^{(1)}_{a,j}- \abrk{{u}^{(1)}_{a}}}^2\,,\\
   &   \abrk{{u}^{(1)}_{a}}= \frac{1}{N}\sum_{j=1}^{N}{u}^{(1)}_{a,j}~.
\end{align}
Here $\sigma_a^2$ and $\abrk{{u}^{(1)}_{a}}$ are the variance and mean of ${u}^{(1)}_{a,j}$ respectively for $N_c=24$ Rb atomic clocks.

Further, the network weighted average pseudo-frequency as calculated using Eq.~(\ref{Eq:Network Wt Avg Pseudofreq}), is partitioned into $M$ windows with each window consisting of $N_w=N/M$ points. We define $\bar{d}^{(1)}_{w,k}\equiv \bar {d}^{(1)}_j$,  where $w=\overline{1,M}$ enumerates the windows and index $k=\overline{1,N_w}$ here spans points in the window $w$, i.e. $j=\prn{w-1}N_w+k$. The standard deviation in a window is calculated as
\begin{align}
    \sigma_w=\sqrt{\frac{1}{N_w-1}\sum_{k=1}^{N_{w}}\prn{\bar{d}^{(1)}_{w,k}-\abrk{\bar{d}_w^{(1)}}}^2}\,,
\end{align}
with
\begin{align}
   &  \abrk{ \bar{d}_w^{(1)}}= \frac{1}{N_w}\sum_{k=1}^{N_w}\bar{d}^{(1)}_{w,k}\,.
\end{align} 

 \begin{figure}[h!]
    \centering
    \includegraphics[width=0.75\linewidth]{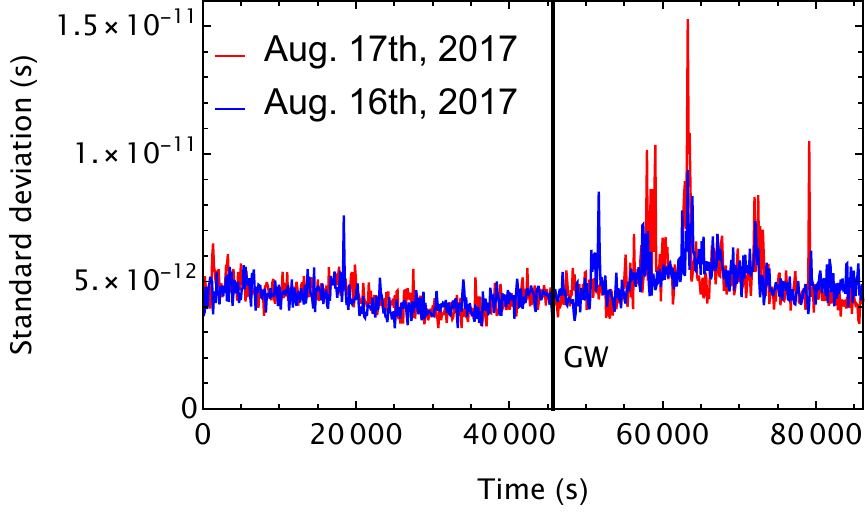} 
    \caption{Standard deviation in a window, $\sigma_w$, of pseudo-frequency for the GPS network average of Rb clocks for August 16 and 17, 2017. GW trigger detected at 45,682~s from the start of the day of August 17, 2017 (GPST), is shown with the black line. In computing standard deviation, we partitioned each day dataset into $M=712$ windows, {each containing} $N_w=121$ data points.}\label{Fig:Std-Dev}
\end{figure}

\begin{figure}[h!]
    \centering
    \includegraphics[width=0.75\linewidth]{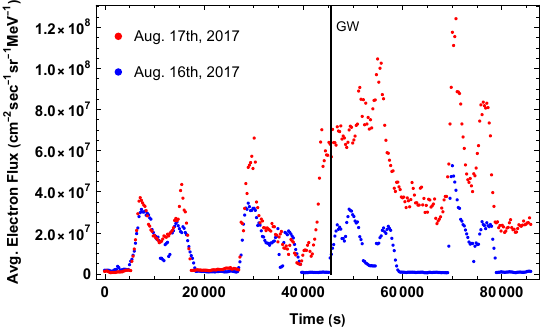} 
    \caption{The average electron flux depicts a period of elevated electron flux that occurs after the GW event, GW170817. The GW LIGO event occurred at 45,682~s on August 17, 2017 (GPST) and is marked with a black vertical line. The elevated electron flux continues onto the following two days, maintaining the same order of magnitude.}
\label{Fig:Electron-Flux}
\end{figure}

In Fig.~\ref{Fig:Std-Dev}, the standard deviation $\sigma_w$ is plotted with respect to time $t_w$ associated with the beginning of the $w^{\text{th}}$ window, $t_w=(w-1)N_w\Delta_t$. The time is based on the GPS Time (GPST) scale offset at the beginning of each GPST day. Note that the difference between GPST and Coordinated Universal Time (UTC) time scale is 18~s for the year 2017. There are two traces: for the day before the GW170817 LIGO trigger (August 16, 2017) and for the day of the event (August 17, 2017). Comparing the two days, we see that the clock network exhibits excess noise level post the GW trigger shown by a black vertical line. In particular, we observe several spikes in the standard deviation of $\bar {d}^{(1)}_j$ occuring after the GW LIGO trigger (on August 17, 2017). The spike with the highest magnitude is delayed by 17,600~s ($\approx 4.9$ hours), from the GW LIGO trigger. Although we see the spike with a high magnitude in both the days' data, the spike post GW trigger is larger by a factor of about 2.3. This factor was computed by taking the average standard deviation for both  days (at $\approx {4.7} \times 10^{-12}\, \mathrm{s}$)  as the floor.  Generically, one would expect both spikes of approximately the same magnitude due to repeating multipath errors in geodetic receivers. It is plausible that this extra noise is due to the putative ELF signal. To quantify this statement additional statistical analysis is required, in particular by finding the optimal size of the window $N_w$ and characterizing the clock statistics. Notice, however, that we found that the solar electron flux was elevated during this period, see Fig.~\ref{Fig:Electron-Flux}, perhaps negating the ELF hypothesis. This is discussed in the following section. 

\section{Systematic effects due to solar activity}
\label{Sec:Solar}
The expulsion of solar energetic particles into the near-Earth space environment during periods of enhanced solar activity can be responsible for compromising GPS radio communication and positioning accuracy. Enhanced solar activity is defined as periods of time in which the solar corona exhibits increased magnetic field deformation. The acceleration of energetic particles from the solar corona, with the total energy release of about $10^{24}$ to $10^{32}$ ergs, typically characterizes a solar flare. Coronal mass ejections (CME) often accompany solar flares, where charged plasma particles with an average velocity ranging from $500$ to $800$ km/s are ejected into the environment ~\cite{Duffin}.

Energetic particles emitted during solar activity events are introduced to Earth's magnetosphere and can precipitate to the ionosphere through acceleration  along magnetic field lines. The energetic electrons and protons can also be trapped within the Van Allen radiation belts, accessible to GPS satellites. Exposure to high energetic particle fluxes within the radiation belts can induce malfunctions to the satellites' electric power system~\cite{Chen}. The precipitation of energetic particles to the ionosphere can introduce ionization density variations, resulting in fluctuations to the amplitude and phase of transmitted signals. This phenomenon is known as ionospheric scintillation. GPS transmitted signals are vulnerable to strong ionospheric scintillation associated with enhanced electron content ~\cite{Kintner}.

The Los Alamos National Laboratory has released energetic particle data recorded by the Combined X-Ray Dosimeter (CXD) aboard 21 operational GPS satellites. The CXD is equipped with high-energy and low-energy particle sensors that can record electron fluxes with an energy of 1 MeV to above 5 MeV and proton fluxes with an energy of 16 MeV to above 60 MeV ~\cite{distel1999}.
An increased sunspot number associated with enhanced solar activity can correspond to elevated energetic particles introduced to the GPS orbit environment. Increased energetic particle exposure can be represented by spikes in electron and proton flux measurements. The daily sunspot number for August of 2017 indicates elevated values occurring at the end of the month. 

We examined the network average proton and electron fluxes from the CXD aboard 17 GPS satellites for 16 August and 17 August, 2017, c.f. Fig.~\ref{Fig:Std-Dev}. The plot of the average electron flux is shown in Fig.~\ref{Fig:Electron-Flux}. We observe a period of elevated electron flux that occurs after the GW event, GW170817. The proton flux exhibited no obvious spike or irregularity before this GW event. The spike in the network weighted average of the pseudo-frequency depicted in Fig.~\ref{Fig:Std-Dev} could be related to elevated energetic particle exposure. Notice that the solar electron flux leads to a charge build-up on a satellite that could randomly discharge after reaching a certain level~\cite{Frederickson}, which could potentially cause frequency jumps in the quartz oscillator~\cite{Riley1981,Dupuis2008}, modifying GPS microwave signals. This elevated electron flux starts around the time of the LIGO trigger. One of the intriguing explanations might be that the GWs perturb the Sun causing the solar activity.

\section{Conclusions}\label{Sec:Conclusion}
The ELF proposal~\cite{dailey2020ELF.Concept} has introduced a novel, exotic physics, modality in multi-messenger astronomy. In this report we outlined some practical considerations for the ELF search with the network of atomic clocks on board GPS satellites. We have generated 1~Hz clock bias data specifically for this analysis. Our search method focused on the excess noise in the clock network weighted average pseudo-frequency. There is an intriguing excess in the clock noise post LIGO gravitational wave trigger for GW170817 binary neutron star merger of August 17, 2017. Potentially it could be explained away by the increased solar electron flux during the observed spike in the clock noise.  It is also intriguing that solar activity is greatly enhanced around the time and after the LIGO trigger.

\section*{Acknowledgments}
This work was supported in part by the U.S. National Science Foundation grant PHY-2207546, by the Heising-Simons Foundation, and by the NASA  Grant 80NSSC23M0104.

\section*{References} 
\bibliographystyle{iopart-num} 

\providecommand{\newblock}{}
\begin{thebibliography}{}
\expandafter\ifx\csname url\endcsname\relax
  \def\url#1{{\tt #1}}\fi
\expandafter\ifx\csname urlprefix\endcsname\relax\def\urlprefix{URL }\fi
\providecommand{\eprint}[2][]{\url{#2}}

\end{thebibliography}


\begin{thebibliography}{10}
	\expandafter\ifx\csname url\endcsname\relax
	\def\url#1{{\tt #1}}\fi
	\expandafter\ifx\csname urlprefix\endcsname\relax\def\urlprefix{URL }\fi
	\providecommand{\eprint}[2][]{\url{#2}}
	
	\bibitem{LIGOfirstObservation2016}
	B~Abbott {\it et al} (2016) {\em Physical Review Letters\/} {\bf 116} 061102 ISSN 0031-9007 \urlprefix\url{https://link.aps.org/doi/10.1103/PhysRevLett.116.061102}
	
	\bibitem{LIGOVirgo-NeutronStarMerger2017}
	B~Abbott {\it et al} (2017) {\em Phys. Rev. Lett.\/} {\bf 119} 161101 ISSN 0031-9007 (\textit{Preprint} \eprint{1710.05836}) \urlprefix\url{https://link.aps.org/doi/10.1103/PhysRevLett.119.161101}
	
	\bibitem{Abbott_multimessenger_2017}
	B~Abbott {\it et al} (2017) {\em The Astrophysical Journal Letters\/} {\bf 848} L13 \urlprefix\url{https://dx.doi.org/10.3847/2041-8213/aa920c}
	
	\bibitem{dailey2020ELF.Concept}
	C~Dailey {\it et al} (2021) {\em Nature Astronomy\/} {\bf 5} 150--158 ISSN 2397-3366 (\textit{Preprint} \eprint{2002.04352}) \urlprefix\url{http://www.nature.com/articles/s41550-020-01242-7}
	
	\bibitem{Derevianko2023-Moriond.ELF}
	Derevianko A (2023)  (\textit{Preprint} \eprint{2305.17138}) \urlprefix\url{http://arxiv.org/abs/2305.17138}
	
	\bibitem{bini2017deviation}
	Bini D, Geralico A and Ortolan A (2017) {\em Phys. Rev. D\/} {\bf 95}(10) 104044 \urlprefix\url{https://link.aps.org/doi/10.1103/PhysRevD.95.104044}
	
	\bibitem{Baumann2019}
	D~Baumann {\it et al} (2019) {\em Phys. Rev. D\/} {\bf 99} 044001 ISSN 2470-0010 \urlprefix\url{https://doi.org/10.1103/PhysRevD.99.044001 https://link.aps.org/doi/10.1103/PhysRevD.99.044001}
	
	\bibitem{Loeb2018-BH-signularities}
	Loeb A (2018)  (\textit{Preprint} \eprint{1805.05865}) \urlprefix\url{https://arxiv.org/abs/1805.05865}
	
	\bibitem{Campbell2021}
	W~Campbell {\it et al} 2021 {\em Phys. Rev. Lett.\/} {\bf 126}(7) 071301 \urlprefix\url{https://link.aps.org/doi/10.1103/PhysRevLett.126.071301}
	
	\bibitem{FlambaumEtAl2004}
	V~Flambaum {\it et al} 2004 {\em Phys. Rev. D\/} {\bf 69}(11) 115006 \urlprefix\url{https://link.aps.org/doi/10.1103/PhysRevD.69.115006}
	
	\bibitem{Roberts2018a}
	B~Roberts {\it et al} 2018 {\em Phys. Rev. D\/} {\bf 97} 083009 ISSN 2470-0010 \urlprefix\url{http://arxiv.org/abs/1803.10264 https://link.aps.org/doi/10.1103/PhysRevD.97.083009}
	
	\bibitem{DanzeyRiley}
	Danzy F and Riley W 1989 Stability test results for gps rubidium clocks {\em Proceedings of the Annual Precise Time and Time Interval ({PTTI}) applications and Planning Meeting (21st), Held in Redondo Beach, California on November 28-30, 1989\/} \urlprefix\url{https://apps.dtic.mil/sti/pdfs/ADA224769.pdf}
	
	\bibitem{Blewitt2015307}
	Blewitt G (2015) {GPS and space-based geodetic methods} {\em Treatise on Geophysics\/} (Oxford: Elsevier) pp 307--338 \urlprefix\url{http://linkinghub.elsevier.com/retrieve/pii/B9780444538024000609}
	
	\bibitem{Riley1981}
	Riley W (1981) A rubidium clock for {GPS} {\em Proceedings of the 13th Annual Precise Time and Time Interval Systems and Applications Meeting\/} \urlprefix\url{https://apps.dtic.mil/sti/pdfs/ADA494259.pdf}
	
	\bibitem{Camparo}
	Camparo J 2007 {\em Physics Today\/} {\bf 60} 33--39 ISSN 0031-9228 (\textit{Preprint} \eprint{https://pubs.aip.org/physicstoday/article-pdf/60/11/33/16696572/33\_1\_online.pdf}) \urlprefix\url{https://doi.org/10.1063/1.2812121}
	
	\bibitem{Guo}
	Y~Guo {\it et al} (2022) {\em AIP Advances\/} {\bf 12} 045216 ISSN 2158-3226 \urlprefix\url{https://doi.org/10.1063/5.0086523}
	
	\bibitem{GriggsEtAl2015}
	Griggs E, Kursinski E~R and Akos D 2015 {\em Radio Science\/} {\bf 50} 813--826 (\textit{Preprint} \eprint{https://agupubs.onlinelibrary.wiley.com/doi/pdf/10.1002/2015RS005667}) \urlprefix\url{https://agupubs.onlinelibrary.wiley.com/doi/abs/10.1002/2015RS005667}
	
	\bibitem{Dupuis2008}
	Dupuis~et al R 2008 {\em Frequency Control Symposium, 2008 IEEE International\/}  655--660
	
	\bibitem{IGS_KOUR}
	International {GNSS} {S}ervice ({IGS}) {F}ile {A}ccess \urlprefix\url{https://files.igs.org/pub/station/log/kour_20230405.log}
	
	\bibitem{ITRF_KOUR}
	{I}nternational {T}errestrial {R}eference {F}rame, ({ITRF}) \urlprefix\url{https://itrf.ign.fr/en/station/KOUR-97301M210}
	
	\bibitem{Major_Book}
	Major F~G 2007 {\em {The Quantum Beat:Principles and Applications of Atomic Clocks; 2nd ed.}\/} (Springer-Verlag) \urlprefix\url{"https://link.springer.com/book/10.1007/978-1-4757-2923-8"}
	
	\bibitem{Klepner1962}
	Kleppner D, Goldenberg H~M and Ramsey N~F 1962 {\em Phys. Rev.\/} {\bf 126}(2) 603--615 \urlprefix\url{https://link.aps.org/doi/10.1103/PhysRev.126.603}
	
	\bibitem{Vessot_2005}
	Vessot R~F~C 2005 {\em Metrologia\/} {\bf 42} S80 \urlprefix\url{https://dx.doi.org/10.1088/0026-1394/42/3/S09}
	
	\bibitem{Lombardi}
	Lombardi M 2002 {\em Time and Frequency\/} pp 783--801 \urlprefix\url{https://www.researchgate.net/publication/269420574_Time_and_Frequency}
	
	\bibitem{Hu}
	Hu W, Shuai T, Xie Y, Chen P, Pei Y, Zhao Y and Wang R 2023 {\em Sensors\/} {\bf 23} ISSN 1424-8220 \urlprefix\url{https://www.mdpi.com/1424-8220/23/22/9202}
	
	\bibitem{Audoin1981}
	{Audoin, C} 1981 {\em Rev. Phys. Appl. (Paris)\/} {\bf 16} 125--130 \urlprefix\url{https://doi.org/10.1051/rphysap:01981001603012500, https://hal.science/jpa-00244902/document}
	
	\bibitem{BERTIGER2020469}
	W~Bertiger {\it et al} 2020 {\em Advances in Space Research\/} {\bf 66} 469--489 ISSN 0273-1177 \urlprefix\url{https://www.sciencedirect.com/science/article/pii/S0273117720302532}
	
	\bibitem{Blewitt_1989}
	Blewitt G (1989) {\em Journal of Geophysical Research: Solid Earth\/} {\bf 94} 10187--10203 (\textit{Preprint} \eprint{https://agupubs.onlinelibrary.wiley.com/doi/pdf/10.1029/JB094iB08p10187}) \urlprefix\url{https://agupubs.onlinelibrary.wiley.com/doi/abs/10.1029/JB094iB08p10187}
	
	\bibitem{Blewitt_1990}
	Blewitt G (1990) {\em Geophysical Research Letters\/} {\bf 17} 199--202 (\textit{Preprint} \eprint{https://agupubs.onlinelibrary.wiley.com/doi/pdf/10.1029/GL017i003p00199}) \urlprefix\url{https://agupubs.onlinelibrary.wiley.com/doi/abs/10.1029/GL017i003p00199}
	
	\bibitem{Bertiger2010}
	Bertiger~et al W 2010 {\em Journal of Geodesy\/} {\bf 84} 327--337 \urlprefix\url{https://doi.org/10.1007/s00190-010-0371-9}
	
	\bibitem{Bohm-GPT2}
	J~B{\"o}hm {\it et al} 2015 {\em GPS Solutions\/} {\bf 19} 433--441 \urlprefix\url{https://doi.org/10.1007/s10291-014-0403-7}
	
	\bibitem{Agnew-repeat}
	D~Agnew {\it et al} 2007 {\em GPS Solutions\/} {\bf 11} 71--76 \urlprefix\url{https://doi.org/10.1007/s10291-006-0038-4}
	
	\bibitem{Larson-repeat}
	K~Larson {\it et al} 2007 {\em Journal of Geophysical Research: Solid Earth\/} {\bf 112} (\textit{Preprint} \eprint{https://agupubs.onlinelibrary.wiley.com/doi/pdf/10.1029/2006JB004367}) \urlprefix\url{https://agupubs.onlinelibrary.wiley.com/doi/abs/10.1029/2006JB004367}
	
	\bibitem{0.A.Howe0.W.Allan1981}
	D~Howe {\it et al} 1981 {Properties of Signal Sources and measurement methods} {\em Proceedings of the 35th Annual Symposium on Frequency Control\/} \urlprefix\url{https://tf.nist.gov/general/pdf/554.pdf}
	
	\bibitem{Riley}
	Riley W and Howe D 2008 Handbook of frequency stability analysis \urlprefix\url{https://tsapps.nist.gov/publication/get_pdf.cfm?pub_id=50505}
	
	\bibitem{Roberts2017-GPS-DM}
	B~Roberts {\it et al} (2017) {\em Nature Comm.\/} {\bf 8} 1195 ISSN 2041-1723 \urlprefix\url{http://www.nature.com/articles/s41467-017-01440-4}
	
	\bibitem{Duffin}
	Duffin R~T, White S~M, Ray P~S and Kaiser M~L 2015 {\em Journal of Physics: Conference Series\/} {\bf 642} 012006 \urlprefix\url{https://dx.doi.org/10.1088/1742-6596/642/1/012006}
	
	\bibitem{Chen}
	Chen Y, Carver M~R, Morley S~K and Hoover A~S 2021 Determining ionizing doses in medium earth orbits using long-term gps particle measurements {\em 2021 IEEE Aerospace Conference (50100)\/} pp 1--21 \urlprefix\url{https://ieeexplore.ieee.org/document/9438516}
	
	\bibitem{Kintner}
	Kintner P~M, Ledvina B~M and de~Paula E~R 2007 {\em Space Weather\/} {\bf 5} (\textit{Preprint} \eprint{https://agupubs.onlinelibrary.wiley.com/doi/pdf/10.1029/2006SW000260}) \urlprefix\url{https://agupubs.onlinelibrary.wiley.com/doi/abs/10.1029/2006SW000260}
	
	\bibitem{distel1999}
	J~Distel {\it et al} 1999 The combined {X}-ray dosimeter ({CXD}) on {GPS} {B}lock {IIR} satellites Tech. rep. Tech. Rep. LA-UR-99 \urlprefix\url{https://permalink.lanl.gov/object/tr?what=info:lanl-repo/lareport/LA-UR-99-2280}
	
	\bibitem{Frederickson}
	Frederickson A, Mullen E, Brautigam D, Kerns K, Robinson P and Holman E 1991 {\em IEEE Transactions on Nuclear Science\/} {\bf 38} 1614--1621 \urlprefix\url{https://ieeexplore.ieee.org/document/124153}
	
\end{thebibliography}
\providecommand{\newblock}{}

\end{document}